# ORIENTATIONAL EFFECTS IN ALKANONE, ALKANAL OR DIALKYL CARBONATE + ALKANE MIXTURES AND IN ALKANONE + ALKANONE OR + DIALKYL CARBONATE SYSTEMS.

FERNANDO HEVIA[(1)], JUAN ANTONIO GONZALEZ[(1)*], CRISTINA ALONSO-TRISTÁN[(2)], ISAÍAS GARCÍA DE LA FUENTE[(1)] AND LUIS FELIPE SANZ[(1)]

[(1)] G.E.T.E.F., Departamento de Física Aplicada, Facultad de Ciencias, Universidad de Valladolid. Paseo de Belén, 7, 47011 Valladolid, Spain.

[(2)] Departamento de Ingeniería Electromecánica. Escuela Politécnica Superior, Universidad de Burgos. Avda. Cantabria s/n. 09006 Burgos, Spain.

*CORRESPONDING AUTHOR, e-mail: jagl@termo.uva.es; Fax: +34-983-423136; Tel: +34-983-423757

**ABSTRACT**

Interactions and structure of alkanone, or alkanal or dialkyl carbonate + alkane mixtures, or of 2-alkanone+ 2-alkanone, or of ketone + dialkyl carbonate systems have been investigated by means of a set of thermodynamic properties and by the application of the Flory model. The properties considered are excess molar quantities: enthalpies, $H_m^E$, volumes, $V_m^E$, or isobaric heat capacities, $C_{pm}^E$, and liquid-liquid equilibria. Experimental data show that alkane mixtures are characterized by rather strong dipolar interactions. In the case of systems containing ketones with the same number of C atoms and a given alkane, dipolar interactions become weaker in the sequence: aromatic > cyclic > linear. In addition, the mentioned interactions become also weaker in the order: dialkyl carbonate > linear alkanone > linear alkanal. This is an important result, as carbonates show lower effective dipole moments than the other compounds, and it suggests that the group size may be relevant when evaluating thermodynamic properties of liquid mixtures. Results on $H_m^E$ from the Flory model show that orientational effects (i.e., non-random mixing) are rather similar for systems with linear, cyclic or aromatic ketones or alkanals and alkanes. In contrast, orientational effects become weaker in dialkyl carbonate + alkane mixtures. The behaviour of 2-alkanone + 2-alkanone systems and of mixtures of longer 2-alkanones or cyclohexanone with dialkyl carbonate is close to random mixing. Larger orientational effects are encountered in solutions of carbonates and shorter 2-alkanones.

# 1. Introduction

We are engaged in a systematic investigation on orientational effects (i.e., non-random mixing) in mixtures of organic liquids by means of the application of different models, such as: DISQUAC [1], ERAS [2], Flory [3], Kirkwood-Buff integrals [4, 5] or the $S_{CC}(0)$ (concentration-concentration structure factor) formalism [6]. In this framework, along a series of works, we have shown that the Flory model is suitable to gain insight into orientational effects present in systems of the type: 1-alkanol + linear alkanone [7], or + nitrile [8], or + linear or cyclic ether [9, 10]; 1-butanol + alkoxyethanol [11], alkoxyethanol + dibutyl ether [11], oxaalkane + alkane [12], or + aromatic compound [13]. Now, we extend these previous studies to alkanone, or alkanal or linear organic carbonate + alkane mixtures. This allows to examine a number of interesting effects in terms of the Flory model: steric effects, cyclization, aromaticity or the group size, which is extremely large in the case of the carbonate group. The latter is rather important, as if the group is too large with respect to the average intermolecular distances, the interaction potential involved could be so complex that no theory would be able to describe it conveniently. This should lead to a poor description of the thermodynamic properties by means of the selected theory. The study is completed with the corresponding treatment of $n$-alkanone + $n$-alkanone, or + linear organic carbonate systems.

Alkane mixtures with linear [14], cyclic [14] or aromatic alkanones [15, 16], or with linear [17] or aromatic alkanals [16, 18], or including linear organic carbonates [19] have been successfully treated in terms of the DISQUAC model. UNIFAC interaction parameters for the contacts CO/CH$_2$ [20] CHO/CH$_2$ [20] or OCOO/CH$_2$ [21] are available in the literature. Interestingly, $n$-alkanone [22, 23], or $n$-alkanal [24] + alkane mixtures have also been studied using only the quasi-chemical approximation of DISQUAC with the coordination number equal to 10, obtaining rather good results. This suggests that orientational effects are not very relevant in such systems.

# 2. The Flory model

In this section, a summary of the Flory model and some results used in this work are presented. More details can be found in the original works [3, 25-28].

## 2.1. Hypotheses

**2.1.1 Hypotheses for pure liquids**. In the Flory model, a liquid, occupying the volume $V$, is formed by $N$ molecules (of mean volume $v_m = V/N$), each of one is divided into $r$ segments. The mean volume of a segment is denoted $v_s = V/rN = v_m/r$. A segment is an arbitrarily chosen isomeric portion of the molecule; its precise definition is left open and may be adapted to circumstances. The core volume of a molecule is defined as $v_m^* = rv_s^*$, where $v_s^*$ is the core volume of a segment. Each segment is endowed with $s$ contacts. The interactions considered are: (i) An attractive intermolecular interaction between pairs of contacts, with a mean potential energy per pair of the form $-\eta/v_s$, where $\eta$ is a positive constant of the liquid considered. (ii) A repulsive interaction, leading to a free volume term in the partition function [29]. (iii) The effect of the rest of the intramolecular interactions is treated assuming [30] that the $3r$ degrees of freedom of a molecule can be divided into two uncoupled categories, i.e., internal and external, and also that, for fluids with densities of liquids, the intramolecular potentials associated with the latter degrees of freedom merely restrict the degrees of freedom per molecule from $3r$ to an effective number of $3rc$. The constant $c \leq 1$ would take into account the restrictions on the precise location of a segment by its neighbors in the same chain. Some parameters of the model are better replaced by the reduction parameters $p^*$ and $T^*$, defined together with the reduced parameters of the liquid, namely, $\bar{T} = T/T^*$, $\bar{p} = p/p^*$ (where $p$ is the pressure and $T$ is the temperature) and $\bar{v} = v_m/v_m^* = v_s/v_s^* = V_m/V_m^*$. In these relations, $V_m = N_A v_m$ denotes the molar volume of the liquid and $V_m^* = N_A v_m^*$ the core molar volume ($N_A$ stands for Avogadro's constant).

**2.1.2 Hypotheses for binary mixtures**. The components of a binary mixture will be indexed by subscripts $i = 1, 2$. Because the definition of segment is arbitrary, it is convenient to impose that the segments of both components have the same core volume. It is supposed that the number of contacts per molecule of a given component is proportional to the core surface area of the corresponding molecule, assumed spherical. The total number of molecules in the mixture is $N = N_1 + N_2$. The total numbers of segments, contacts and effective number of degrees of freedom ($rN$, $srN$ and $3rcN$) are taken as additive. It is convenient to define the segment and contact fractions, respectively, by $\varphi_i = r_i N_i / rN$ and $\theta_i = s_i r_i N_i / srN = \varphi_i s_i / s$. Of course, $\sum \varphi_i = \sum \theta_i = 1$. It is also assumed that the mean intensity of the interaction between segments of molecules of the same component is the same in the mixture as in the pure species; the total intermolecular energy of the binary mixtures can be written in the same form as for pure compounds, by defining $v_s = V/rN$, and the $\eta$ parameter for the mixture by $\eta = \theta_1 \eta_1 + \theta_2 \eta_2 - A_{12} \Delta \eta / (srN)$. Here, $A_{ii}$ and $A_{12}$ are the numbers of pairs of contacts between equal and different molecules respectively, $\Delta \eta = \eta_1 + \eta_2 - 2\eta_{12}$ and $\eta_{12}$ characterizes the mean intensity of the interaction between segments of different molecules. Moreover, **random mixing** is assumed. This hypothesis states that, given a contact, the remaining contacts in the mixture have the same probability of forming an interacting pair with it. It is expressed by the equations $A_{12,\text{random}} = srN\theta_1\theta_2$ and $\eta_{\text{random}} = \theta_1\eta_1 + \theta_2\eta_2 - \theta_1\theta_2\Delta\eta$.

**2.1.3 Equations**. For both pure compounds and binary mixtures, the intermolecular molar energy, $E_m$, and the thermal equation of state (in reduced form) are:

$$E_m = -\frac{p^* V_m^*}{\bar{v}} = -TV_m^* \frac{p^*}{T^*} \frac{1}{\bar{v}\bar{T}} \tag{1}$$

$$\frac{\bar{p}\bar{v}}{\bar{T}} = \frac{\bar{v}^{1/3}}{\bar{v}^{1/3} - 1} - \frac{1}{\bar{v}\bar{T}} \tag{2}$$

(the last equality of eq. (1) is useful when treating mixtures; see below). The so-called geometrical parameter of the mixture, $S_{12} = s_1/s_2$, is

$$S_{12} = \left(\frac{V_{m1}^*}{V_{m2}^*}\right)^{-1/3} \tag{3}$$

The relation of the parameters of the mixture and of the pure compounds is

$$V_m^* = x_1 V_{m1}^* + x_2 V_{m2}^* \tag{4}$$

$$\varphi_i = \frac{x_i V_{mi}^*}{V_m^*} \tag{5}$$

$$\theta_2 = 1 - \theta_1 = \frac{\varphi_2}{\varphi_2 + S_{12}\varphi_1} \tag{6}$$

$$p^* = \varphi_1 p_1^* + \varphi_2 p_2^* - \varphi_1 \theta_2 X_{12} \tag{7}$$

$$\frac{p^*}{T^*} = \varphi_1 \frac{p_1^*}{T_1^*} + \varphi_2 \frac{p_2^*}{T_2^*} \tag{8}$$

where $x_i$ is the mole fraction of component $i$ and, in eq. (7), the parameter $\Delta\eta$ has been replaced by the so-called energetic parameter $X_{12} = s_1 \Delta\eta / 2(v_s^*)^2$. Also, using eqs. (1) and (2) one can derive [25] simple expressions to obtain $V_m^*$ and $p^*$ of the pure compounds in terms of experimental molar volumes and coefficients of isobaric thermal expansion, $\alpha_p$, and isothermal compressibility, $\kappa_T$:

$$V_m^* = V_m \left[\frac{3T\alpha_p + 3(1 - 2p\kappa_T)}{4T\alpha_p + 3(1 - 2p\kappa_T)}\right]^3 \tag{9}$$

$$p^* = \left(\frac{T\alpha_p}{\kappa_T} - p\right)\bar{v}^2 \tag{10}$$

Ignoring the difference between internal energy and enthalpy in condensed systems at low pressure, the molar excess enthalpy, $H_m^E$, can be calculated from the molar intermolecular energies of the mixture and of the pure compounds, $H_m^E = E_m - x_1 E_{m1} - x_2 E_{m2}$, or:

$$H_m^E = x_1 p_1^* V_{m1}^* \left(\frac{1}{\bar{v}_1} - \frac{1}{\bar{v}}\right) + x_2 p_2^* V_{m2}^* \left(\frac{1}{\bar{v}_2} - \frac{1}{\bar{v}}\right) + \frac{x_1 V_{m1}^* \theta_2 X_{12}}{\bar{v}} \tag{11}$$

The part in eq. (11) containing $X_{12}$ is named the interactional term, $H_{m,int}^E$. The rest of the contributions are called the equation of state term, $H_{m,eos}^E$. The molar volume $V_m = \bar{v} V_m^*$ of the mixture is known from the equation of state, which permits to calculate as well the molar excess volume $V_m^E = V_m - x_1 V_{m1} - x_2 V_{m2}$.

### 2.2. Estimation of the Flory energetic parameter

From the composition, pressure, temperature and the reduction parameters of the pure liquids, there are several quantities that can be directly calculated: $S_{12}, V_m^*, \varphi_i, \theta_i$ and the ratios $p^*/T^*$ and $\bar{p}/\bar{T}$ (see eqs. (3), (4)-(6) and (8)). A procedure to obtain the energetic parameter $X_{12}$ from $H_m^E$ at a given composition without approximations will be now exposed. From $H_m^E$, the value of $E_m$ follows. Next, use the second equality of eq. (1) to obtain ($1/\bar{v}\bar{T}$). Then, solve the equation of state for $\bar{v}$, and use the first equality of eq. (1) to determine $p^*$. Finally, $X_{12}$ can be calculated from eq. (7).

### 2.3. Study of the random mixing hypothesis

If random mixing is not considered but the definition of $X_{12}$ is used, then one can write

$$\eta = \theta_1 \eta_1 + \theta_2 \eta_2 - X_{12} \frac{A_{12}}{srN} \frac{2(v_s^*)^2}{s_1} = \theta_1 \eta_1 + \theta_2 \eta_2 - \theta_1 \theta_2 X_{12}'(x_1) \frac{2(v_s^*)^2}{s_1} \tag{12}$$

where the composition-dependent parameter $X_{12}'(x_1)$ has been defined by

$$X_{12}'(x_1) = X_{12} \frac{A_{12}}{A_{12,random}} \tag{13}$$

When the random mixing hypothesis is excluded, the equations of the model for binary mixtures have the same form as when including it, by replacing $X_{12}$ by $X_{12}'(x_1)$. Therefore, one can estimate $X_{12}'(x_1)$ from $H_m^E$ at different compositions by exactly the same procedure considered before for $X_{12}$ (see 2.2). Furthermore, if $X_{12}$ (the Flory energetic parameter of the mixture) is considered as known, it is possible to study the deviations from the random mixing hypothesis as a function of composition, by means of the quantity $X'_{12}(x_1)/X_{12} = A_{12}/A_{12,\text{random}}$. In this procedure, the selection of a criterion for the value of $X_{12}$ is implicit. Note that if $X_{12}$ is estimated from $H_m^E$ for some $x_1$ value, then the value obtained for $X'_{12}(x_1)/X_{12}$ at that composition will be 1, because, in the estimation of $X_{12}$, random mixing is assumed.

## 3. Results

The physical properties ($V_m$, molar volume; $\alpha_p$, isobaric coefficient of thermal expansion; $\kappa_T$, coefficient of isothermal compressibility) and Flory reduction parameters ($V_m^*$, reduction molar volume; and $p^*$, reduction pressure) of the pure compounds at temperature $T = 298.15$ K and pressure $p = 0.1013$ MPa are listed in **Table S1**. At $T \neq 298.15$ K, the values of the mentioned properties have been estimated using the well-known equations [31]:

$$V_m = V_{m0} \exp(\alpha_{p0} \Delta T) \tag{14}$$

$$\alpha_p = \alpha_{p0} + \alpha_{p0}^2 (7 + 4\alpha_{p0} T) \Delta T / 3 \tag{15}$$

$$(\alpha_p / \kappa_T) = (\alpha_{p0} / \kappa_{T0}) - (\alpha_{p0} / \kappa_{T0})(1 + 2\alpha_{p0} T) \Delta T / T \tag{16}$$

where the subscript 0 refers to the property at 298.15 K and $\Delta T = T - 298.15$ K.

The Flory energetic parameters $X_{12}$ have been estimated from $H_m^E$ values at equimolar composition (**Table 1**). Results on $H_m^E$ and $V_m^E$ obtained from the Flory model using them are compared with the experimental values in **Table 1** and **Table S2** respectively. **Table 1** also includes the different contributions to $H_m^E$ and the relative standard deviations for $H_m^E$, defined as:

$$\sigma_r(H_m^E) = \left[\frac{1}{N}\sum\left(\frac{H_{m,\text{exp}}^E - H_{m,\text{calc}}^E}{H_{m,\text{exp}}^E}\right)^2\right]^{1/2} \tag{17}$$

where the sum is taken for $N = 19$ data points, and $H_{m,\text{exp}}^E$ represents smoothed $H_m^E$ values from those reported in the original works calculated at $\Delta x_1 = 0.05$ in the composition range [0.05, 0.95] by means of Redlich-Kister expansions. The mean values of $\sigma_r(H_m^E)$ for some series of compounds are shown graphically in **Fig. 1** and **Fig. 2**. They are defined as

$$\langle \sigma_r(H_m^E) \rangle = \frac{1}{N_s} \sum \sigma_r(H_m^E) \qquad (18)$$

where $N_s$ is the number of systems considered.

## 4. Discussion

From now on, unless explicitly stated otherwise, we will refer to values of thermodynamic properties at 298.15 K and, in the case of excess functions, at equimolar composition. Also, $n$ will stand for the number of C atoms in the $n$-alkane.

The differences between the intermolecular forces of homomorphic compounds may be estimated by the corresponding difference in their standard molar enthalpies of vaporization at 298.15 K, $\Delta H_{m,v}$. To evaluate the weight of non-dispersive interactions in the compounds under study, the quantity $\Delta\Delta H_{m,v}$ [32-34] has been calculated:

$$\Delta\Delta H_{m,v} = \Delta H_{m,v}(\text{compound}) - \Delta H_{m,v}(\text{homomorphic hydrocarbon}) \qquad (19)$$

Moreover, the effect of polarity in bulk properties can be examined by means of the effective dipole moment, $\bar{\mu}$, defined by [35-38]:

$$\bar{\mu} = \mu \left( \frac{N_A}{4\pi\varepsilon_0 V_m k_B T} \right)^{1/2} \qquad (20)$$

where $N_A$ is Avogadro's constant, $\varepsilon_0$ the vacuum permittivity, $k_B$ Boltzmann's constant, $T$ is the temperature and $\mu$ is the dipole moment. Values of $\Delta\Delta H_{m,v}$ and $\bar{\mu}$ of some pure compounds considered in this work can be found in **Table 2**.

### 4.1. Ketone, aldehyde or linear organic carbonate + alkane systems

The excess molar enthalpies, $H_m^E$, of $n$-alkanone + $n$-alkane systems are large and positive, arising from the disruption of strong dipolar interactions between ketone molecules along the mixing process. On the one hand, for mixtures with a given $n$-alkanone, $H_m^E$ increases with $n$. For example, $H_m^E$(2-butanone)/J·mol$^{-1}$: 1160 ($n$ = 5, [39]) < 1254 ($n$ = 6, [40]) < 1338 ($n$ = 7, [39]) < 1408 ($n$ = 8, [39]) < 1545 ($n$ = 10, [39]) < 1661 ($n$ = 12, [41]) < 1863 ($n$ = 16, [41]). These values may be explained in terms of an increase of the number of ketone-ketone interactions broken by longer $n$-alkanes. On the other hand, $H_m^E$ of systems including a fixed $n$-alkane decreases when the alkanone size is increased. In fact, $H_m^E(n = 7)$/J·mol$^{-1}$: 1676 (2-propanone, [42]) > 1338 (2-butanone, [39]) > 1135 (2-pentanone, [39]) > 1055 (2-hexanone, [43]) > 886 (2-heptanone, [44]). This behavior can be ascribed to a weakening of interactions between molecules of longer alkanones, because their CO group is more sterically hindered. Note that the $\Delta\Delta H_{m,v}$ and $\bar{\mu}$ values of 2-alkanones also decrease when the ketone size increases (**Table 2**).

In addition, $H_m^E$ values of cyclohexane systems are larger than those of mixtures containing hexane (geometrical effect). For example, $H_m^E$(2-hexanone)/J·mol$^{-1}$ = 949 ($n$ = 6, [43]), 1044 (cyclohexane, [45]). The same occurs for many other mixtures, as those involving a linear amine [46] or an oxaalkane [33].

The excess molar volume, $V_m^E$, values of $n$-alkanone + $n$-alkane mixtures are also large and positive and change in line with $H_m^E$. For instance, $V_m^E$ ($n$ = 7)/ cm$^3$·mol$^{-1}$ = 1.130 (2-propanone, [42]) > 0.803 (2-butanone, [47]) > 0.375 (2-hexanone, [48]). Moreover, $V_m^E$ increases with $n$ in mixtures with a given ketone, as it is shown by the following experimental results (in cm$^3$·mol$^{-1}$) for 2-butanone systems: 0.803 ($n$ = 7) < 0.866 ($n$ = 8) [49]) < 0.952 ($n$ = 10, [47]) < 0.996 ($n$ = 12, [47]). It is clear that the main contribution to $V_m^E$ of the mentioned solutions arises from interactional effects.

The existence of strong dipolar interactions in $n$-alkanone mixtures is also supported by their relatively high upper critical solution temperatures (UCST). In the case of 2-propanone systems, UCST/K = 286.2 ($n$ = 12), 290.6 ($n$ = 14); 300.2 ($n$ = 16) [50].

$n$-Alkanone + $n$-alkane systems show rather low $C_{pm}^E$ values (excess molar isobaric heat capacity), which is a typical feature of mixtures characterized by dipolar interactions. For example, $C_{pm}^E$/J·mol$^{-1}$·K$^{-1}$ = 2.9 (2-propanone + $n$-C$_7$, [51]); 3.7 (2-propanone + $n$-C$_{16}$, [52]); 1.8 (2-butanone + $n$-C$_7$, [47]). Their $C_{pm}^E(x_1)$ curves are S-shaped (2-butanone + $n$-C$_7$, [47]) or W-shaped (2-propanone, or 2-butanone + $n$-C$_{12}$, [52]). The latter have been interpreted in terms of non-random effects, typically more important when temperature is close to the UCST. The $C_{pm}^E$ function then depends strongly on temperature (it is a decreasing function of $T$) and on the length of the $n$-alkane [52].

$H_m^E$ values of systems containing a cyclic alkanone (1035 J·mol$^{-1}$ for the cyclohexanone + $n$-C$_6$ mixture [53]) are larger than those of mixtures with homomorphic 2-alkanones (cyclization effect). $\Delta\Delta H_{m,v}$ and $\bar{\mu}$ values of cycloalkanones are also higher. For example, $\Delta\Delta H_{m,v}$/kJ·mol$^{-1}$ and $\bar{\mu}$ values are, respectively, 12.05 [54], 1.337 [55] for cyclohexanone, and 11.41 [56], 0.913 [55] for 2-hexanone. This allows to conclude that ketone-ketone interactions are then stronger in the case of cycloalkanones. Interestingly, $V_\mathrm{m}^\mathrm{E}$ values of mixtures with a given alkane are higher for solutions with 2-alkanones, indicating that structural effects are more relevant in cycloalkanone systems. In this framework, the negative $V_\mathrm{m}^\mathrm{E}$ value of the cyclohexanone + $n$-$C_6$ mixture (–0.323 cm$^3$·mol$^{-1}$ [53]) remarks the existence of strong structural effects in such a solution, as the corresponding $H_\mathrm{m}^\mathrm{E}$ value is positive [57].

Acetophenone mixtures are characterized by rather high UCST/K values: 277.36 ($n$ = 10), 283.57 ($n$ = 12), 289.90 ($n$ = 14 and 295.16 ($n$ = 16) [15], and, consequently, the corresponding $H_\mathrm{m}^\mathrm{E}$ values are also large (1620 J·mol$^{-1}$ [58] for the decane solution). The latter value is considerably higher than that of the 2-octanone + $n$-$C_{12}$ mixture (1068 J·mol$^{-1}$ [59]). Therefore, interactions between alkanone molecules are stronger in systems involving aromatic ketones. This has been attributed to the existence of proximity effects between the CO group and the aromatic ring placed in the same molecule, which leads to enhanced dipolar alkanone-alkanone interactions [15, 16]. Accordingly, values of $\Delta\Delta H_\mathrm{m,v}$ and $\bar{\mu}$ of acetophenone are higher than those of 2-octanone (**Table 2**).

Mixtures of $n$-alkanal or 2-alkanone + $n$-alkane behave similarly. For example, $H_\mathrm{m}^\mathrm{E}$ ($n$ = 7)/ J·mol$^{-1}$ = 1796 > 1457 > 1218 > 1066 for ethanal, propanal, butanal and pentanal respectively [24]. It is noteworthy that: (i) systems formed by 2-alkanone and heptane show higher $H_\mathrm{m}^\mathrm{E}$ values than those of mixtures with homomorphic $n$-alkanals (see above); (ii) values of $\Delta\Delta H_\mathrm{m,v}$ and $\bar{\mu}$ are also higher for 2-alkanones. For instance, $\Delta\Delta H_\mathrm{m,v}$/kJ·mol$^{-1}$ = 16.20 (2-propanone) > 14.83 (propanal) [54], and $\bar{\mu}$ = 1.281 (2-propanone) > 1.214 (propanal) [55]. Thus, intermolecular interactions are stronger in $n$-alkanone systems.

Proximity effects are also present in benzaldehyde mixtures, which show the following UCST/K values: 278.54 ($n$ = 10), 284.74 ($n$ = 12), 290.06 ($n$ = 14) and 297.98 ($n$ = 16) [18]. In addition, $H_\mathrm{m}^\mathrm{E}$(benzaldehyde + $n$-heptane) = 1360 J·mol$^{-1}$ [60] is larger than the value for the pentanal mixture. This shows again that the interactions are stronger when the carbonyl group is attached to an aromatic ring, and it is as well remarked by the corresponding $\Delta\Delta H_\mathrm{m,v}$ and $\bar{\mu}$ values (**Table 2**). On the other hand, UCST values are slightly higher for benzaldehyde mixtures than for acetophenone solutions. It seems that dipolar interactions are slightly stronger when the alkanal is involved.

Comments similar to those given above for 2-alkanone + alkane mixtures are valid for the corresponding systems with linear organic carbonates. The large positive values of $H_\mathrm{m}^\mathrm{E}$ ($n$ = 7)/J·mol$^{-1}$ = 1988 (DMC, [61]); 1328 (DEC, [62]) and of the UCST/K for DMC solutions (297.62, 307.61 and 316.21 for $n$ = 12,14,16 [63] respectively) point out to the existence of strong interactions between carbonate molecules. The $V_m^E$ values are very large and change in line with $H_m^E$. For example, $V_m^E$ (DMC)/cm$^3 \cdot$mol$^{-1}$ = 1.158 ($n$ = 7); 1.442 ($n$ = 10) [64]. Interestingly, $H_m^E$ of mixtures with the same alkane and homomorphic 2-alkanones (2-propanone, 3-pentanone) are lower. All these considerations reveal that dipolar interactions are weaker in ketone systems. In this framework, the very low $\bar{\mu}$ values of DMC (0.391) and DEC (0.312) [65] are remarkable. They suggest that the size of the group may also play an important role when determining the thermodynamic properties.

### 4.2. Ketone + ketone or organic carbonate systems

In 2-alkanone + alkane mixtures, the replacement of the alkane by a ketone or a linear organic carbonate leads to decreased $H_m^E$ values, which underlines the existence of interactions between unlike molecules. Neglecting structural effects [35, 66], it is possible to consider that the $H_m^E$ of A + B mixture (here, A is a ketone and B is either a ketone or a linear organic carbonate) is the result of different contributions. Positive contributions, $\Delta H_m^{A-A}$ and $\Delta H_m^{B-B}$, arise from the breaking of interactions between like molecules during the mixing process, which can be evaluated from $H_m^E$ data for A or B + heptane systems, respectively. The negative contribution, $\Delta H_m^{A-B}$, comes from the creation of interactions between unlike molecules. At equimolar composition, $\Delta H_m^{A-B}$ may be then determined from the expression:

$$\Delta H_m^{A-B} = H_m^E(A+B) - H_m^E(A+\text{heptane}) - H_m^E(B+\text{heptane}) \qquad (21)$$

The calculated values of $\Delta H_m^{A-B}$ are listed in **Table 3**. We note that $\left|\Delta H_m^{A-B}\right|$ diminishes when, for a given solute (e.g., 2-propanone), the solvent size (2-alkanone or carbonate) increases. This clearly indicates that, at such conditions, interactions between unlike molecules are weakened, probably because the polar group of the solvent is more sterically hindered. It explains the observed increase of $H_m^E$ along the homologous series 2-propanone + organic solvent (= 2-alkanone, carbonate), as the $\Delta H_m^{A-B}$ term is less negative. In addition, ketone-carbonate interactions are quite similar to those of ketone-ketone type. For example, $\left|\Delta H_m^{A-B}\right|$/J·mol$^{-1}$ = 2985 (2-propanone + 2-butanone); 3052 (DMC + 2-butanone); 2708 (2-propanone + 2-pentanone); 2747 (DMC + 2-pentanone). Similarly, mixtures with carbonates and 2-hexanone or cyclohexanone show similar $\left|\Delta H_m^{A-B}\right|$ values. The cyclization effect is here of minor importance.

It is remarkable that $H_m^E$ values of 2-alkanone (A) + carbonate (B), systems are higher for solutions with DMC, except for the acetone mixture. This can be analyzed taking into account the relative weight of the different contributions to $H_m^E$: (i) the positive $\Delta H_m^{B-B}$ term is higher for DMC systems; (ii) however, the positive $\Delta H_m^{A-A}$ contribution should be higher for DEC systems, due to the larger aliphatic surface of this carbonate; (iii) the $\Delta H_m^{A-B}$ term is more negative when DMC is involved (**Table 3**). For mixtures not including acetone, $H_m^E$(DMC) > $H_m^E$(DEC), and this indicates that, in DMC mixtures, the increased $\Delta H_m^{B-B}$ value compensates the lower $\Delta H_m^{A-A}$ + $\Delta H_m^{A-B}$ result. For the acetone mixtures, the opposite behavior is encountered and $H_m^E$(DMC) < $H_m^E$(DEC).

**4.3. Results from the Flory model**

For alkane mixtures, $X_{12}$ values are large and positive, indicating the existence of strong interactions between like polar molecules. We note that $H_{m,int}^E$ is the dominant contribution to $H_m^E$ (**Table 1**). Nevertheless, the $H_{m,eos}^E$ term, which depends on the reduced volume of the solution compared to those of the pure components, is rather large. This contribution is ranged between 32.5% for ethanol + heptane and 16.5% for 2-butanone + 2-heptanone (**Table 1**). As a general trend, better results are obtained the lower the $H_{m,eos}^E$ values are. On the other hand, $X_{12}$ and $H_m^E$ change in line (**Table 1**), which remarks the relevance of interactional effects on $H_m^E$. Interestingly, in the case of 2-alkanone or *n*-alkanal + heptane mixtures, a linear relationship exists between $\Delta\Delta H_{vap}$ and $X_{12}$. In fact, $\Delta\Delta H_{vap}$(2-alkanone)/kJ·mol$^{-1}$ = 7.86 + 0.089 $X_{12}$/MPa (*r*, correlation coefficient, 0.994); and $\Delta\Delta H_{vap}$(*n*-alkanal)/kJ·mol$^{-1}$ = 4.84 + 0.0125 $X_{12}$/MPa (*r* = 0.999). This also supports the close relation between $X_{12}$ and interactional effects. We have applied the same approach to dialkyl ether + heptane systems using the values of $\Delta\Delta H_{vap}$ and $X_{12}$ reported previously in our investigation on oxaalkane + alkane systems in terms of the Flory model [12]. The result is: $\Delta\Delta H_{vap}$/kJ·mol$^{-1}$ = − 2.243 + 0.1814 $X_{12}$/MPa (*r* = 0.982). We note that for not very large $X_{12}$ values (e.g., 7.10 MPa for the dipropyl ether mixture [12]), $\Delta\Delta H_{vap}$ values become negative, which reveals very weak interactions between dialkyl ether molecules). The replacement of an alkane by an *n*-alkanone or a linear organic carbonate leads to decreased $X_{12}$ values, due to the creation of interactions between unlike molecules during the mixing process (**Table 1** and **Table 3**). For 2-propanone + 2-alkanone mixtures, the linear dependence between $\Delta\Delta H_{vap}$ and $X_{12}$ is somewhat poorer: $\Delta\Delta H_{vap}$/kJ·mol$^{-1}$ = 13.2 − 0.144 $X_{12}$/MPa (*r* = 0.958). Here, the negative slope deserves attention, as it implies that when $\Delta\Delta H_{vap}$ increases $X_{12}$ decreases, which means that interactions between unlike molecules become more relevant.

Inspection of **Table S2** shows that the theoretical $V_\text{m}^\text{E}$ results are larger than the experimental values, as the interactional contribution from the Flory model to this excess function is overestimated. Nevertheless, the relative variation of $V_\text{m}^\text{E}$ with the alkane size in mixtures with a given polar component is correctly given by the model. The same trend is also encountered, for example, in oxaalkane + alkane mixtures [12]. The $V_\text{m}^\text{E}$ results can be better rationalized by means of the Prigogine-Flory-Patterson model [67]. Here, $V_\text{m}^\text{E}$ is written as the sum of three contributions: an interactional contribution, a curvature term and the so-called $p^*$ term. The second one depends on $-(\bar{v}_1 - \bar{v}_2)^2$ and is always negative. The last one depends on $(p_1^* - p_2^*)(\bar{v}_1 - \bar{v}_2)$. For the alkanone + alkane systems considered, $p_1^* > p_2^*$ and the sign of the $p^*$ term depends on that of $(\bar{v}_1 - \bar{v}_2)$. For the cyclohexanone + hexane, or + heptane mixtures, the $(\bar{v}_1 - \bar{v}_2)$ values are negative and rather large. The contributions from the curvature and $p^*$ terms are then also large and negative and the model predicts low $V_\text{m}^\text{E}$ values for these systems. Particularly, $V_\text{m}^\text{E}$ (hexane)/cm$^3$·mol$^{-1}$ = –0.323 (experimental); –0.194 (calculated). For $(\bar{v}_1 - \bar{v}_2) > 0$, the $p^*$ contribution is also positive and may be rather large. The corresponding $V_\text{m}^\text{E}$ values are then largely overestimated, as the curvature term is usually small in absolute value.

### 4.3.1. Alkanone + alkane

We note that, in the framework of the Flory model, orientational effects are quite similar in 2-alkanone mixtures. It is noteworthy that the theoretical results are somewhat poorer when temperature is not far from the corresponding UCST. This is the case of the systems 2-propanone + hexane at 243.15 K (UCST = 237.2 K [68]; $\sigma_r(H_\text{m}^\text{E}) = 0.183$); or + decane at 298.15 K (UCST = 266.8 K [69]; $\sigma_r(H_\text{m}^\text{E}) = 0.249$); or + hexadecane at 298.15 K (UCST = 300.6 K [41]; $\sigma_r(H_\text{m}^\text{E}) = 0.179$). The mixtures 2-butanone + dodecane ($\sigma_r(H_\text{m}^\text{E}) = 0.214$), or + tetradecane ($\sigma_r(H_\text{m}^\text{E}) = 0.214$) at 298.15 K are also close to the UCST and the $C_{p\text{m}}^\text{E}$ curves are W-shaped [47], a behavior typically ascribed to non-random effects. Interestingly, the position of the CO group has not a special relevance on orientational effects. In fact, $\langle \sigma_r(H_\text{m}^\text{E}) \rangle = 0.144$ (2-pentanone) ≈ 0.140 (3-pentanone). In addition, $\langle \sigma_r(H_\text{m}^\text{E}) \rangle$ values are also rather similar for homologous series including linear or cyclic alkanones. In fact, $\langle \sigma_r(H_\text{m}^\text{E}) \rangle$ = 0.144 (2-pentanone); 0.156 (cyclopentanone); 0.152 (2-hexanone); 0.136 (cyclohexanone). We have applied the UNIFAC model (Dortmund version) [43] to linear or cyclic alkanone + alkane mixtures assuming that these systems differ merely by the presence of the *c*-CH$_2$ group in cyclic

alkanones –in other words, using the same set of interaction parameters [43] for those contacts where the CO group participates. The results are: $\langle \sigma_r(H_m^E) \rangle = 0.063$ (linear alkanone; $N_S = 42$), and 0.086 (cyclic alkanone; $N_S = 13$). It is remarkable that the largest differences for cycloalkanone mixtures are encountered for the systems with hexadecane ($\sigma_r(H_m^E) = 0.151$ for the cyclopentanone solution). This may be ascribed, at least to some extent, to the existence of the Patterson effect [70, 71]. It is known that this effect leads to an extra endothermic contribution to $H_m^E$, attributed to the destruction of correlations of molecular orientations existing between long $n$-alkanes by a globular or a plate-like molecule [70, 71]. The UNIFAC results reveal that: (i) orientational effects are practically independent on the $n$-alkanone under consideration; (ii) the cyclization effect is rather negligible. In the case of acetophenone mixtures, the Flory model provides $\langle \sigma_r(H_m^E) \rangle = 0.135$, a close value to that of 2-octanone systems (0.142). Nevertheless, $\sigma_r(H_m^E)$ is somewhat large for the acetophenone + decane mixture (0.151), whose UCST = 277.4 K [15].

Of course, the rise in temperature is linked in most cases with better theoretical results (**Table 1**), which agrees with the fact that random mixing effects are more relevant with this increase. For example, for the 2-propanone + $n$-hexane system, $\sigma_r(H_m^E) = 0.183$ (243.15 K) > 0.151 (293.15 K) [72].

**4.3.2. Alkanal + alkane**

The larger $\sigma_r(H_m^E)$ values obtained for the propanal + hexane (0.226), or + heptane (0.333) mixtures are remarkable, as they do not fit into the results for the remaining systems. These large values may be due to experimental inaccuracies. Or they might arise because some of the $H_m^E$ values at low mole fractions of one component are not reliable enough, since they were obtained from Redlich-Kister regressions with coefficients determined from $H_m^E$ measurements at the central range of composition. Similar comments are also valid for benzaldehyde mixtures. Nevertheless, Flory results are very close for butanal or 2-butanone mixtures. Orientational effects do not change when replacing the CO group by the CHO group.

**Fig. 3** shows the plots of $X'_{12}(x_1)/X_{12}$ for several systems containing heptane as functions of $x_1$. It must be remarked that the shape of this function is, to a large extent, not dependent on the 2-alkanone considered, or on if the 2-alkanone is replaced by a cycloalkanone, acetophenone or an $n$-alkanal. This supports that orientational effects are similar for all the systems studied with the CO or the CHO groups. In contrast, such a general behavior is not observed in ether + heptane systems [12], as it is shown in **Fig. 4** for various systems of this type realized using data reported earlier [12].

### 4.3.3. Linear organic carbonate + alkane

Experimental data show that interactions between polar linear molecules become stronger in the sequence: alkanal < alkanone < carbonate. As already pointed out, $H_m^E$ and UCST values are higher for carbonate mixtures. However, theoretical results are slightly better for systems with DMC or DEC than for solutions with 2-propanone or 3-pentanone, indicating that orientational effects are weaker in carbonate systems. This might be explained in terms of higher $TS_m^E$ values ($TS_m^E = H_m^E - G_m^E$; $G_m^E$, excess molar Gibbs energy) for carbonate mixtures. The application of the DISQUAC model to $n$-alkanone [37] or linear carbonate [19] + heptane mixtures at 298.15 K provides $G_m^E$/J·mol$^{-1}$ = 1112 (2-propanone); 715 (3-pentanone); 1199 (DMC); 788 (DEC). Using these values, we obtain $TS_m^E$ (heptane)/J·mol$^{-1}$ = 789 (DMC); 564 (2-propanone); 540 (DEC); 363 (3-pentanone).

### 4.3.4. $n$-alkanone + $n$-alkanone, or + alkanone + linear organic carbonate

For 2-alkanone + 2-alkanone systems (**Fig. 2**), the random mixing hypothesis is valid to a rather large extent. The replacement of the CO group by the OCOO group has a negative impact on the results from the model. For 2-alkanone + dialkyl carbonate systems, $\langle \sigma_r(H_m^E) \rangle$ = 0.293 (2-propanone), 0.343 (2-butanone), 0.282 (2-pentanone), 0.235 (2-hexanone), 0.109 (2-octanone), and 0.044 (2-undecanone). These results underline that ketone-carbonate interactions are of a more polar nature for the shorter 2-alkanones, as it is seen as well from the $\Delta H_m^{A-B}$ values (**Table 3**). In contrast, the behavior of the mixtures cyclohexanone + diakyl carbonate is close to random mixing.

## 5. Conclusions

The Flory model has been applied to alkanone, or alkanal or linear organic carbonate + alkane mixtures, and to $n$-alkanone + $n$-alkanone or alkanone + dialkyl carbonate systems. For alkane mixtures, the mean relative standard deviation for the excess molar enthalpies, $\langle \sigma_r(H_m^E) \rangle$, is: 0.152 ($n$-alkanone); 0.134 (cyclic alkanone); 0.135 (acetophenone); 0.192 ($n$-alkanal); 0.213 (benzaldehyde); 0.107 (linear organic carbonate). These results show that orientational effects in alkane mixtures are practically independent of steric effects, or of the position of the polar group in a linear chain, in a ring or in an aromatic ring. In addition, orientational effects are similar in systems containing alkanones or alkanals. Binary systems formed by two $n$-alkanones show a behaviour close to random mixing ($\langle \sigma_r(H_m^E) \rangle$ = 0.039). In contrast, orientational effects are strong in systems with shorter $n$-alkanones and dialkyl carbonates ($\langle \sigma_r(H_m^E) \rangle = 0.230$). Mixtures with longer alkanones and cyclohexanone are close to random mixing ($\langle \sigma_r(H_m^E) \rangle = 0.047$).

## Funding

The authors gratefully acknowledge the financial support received from the Consejería de Educación y Cultura of Junta de Castilla y León, under Project BU034U16. F. Hevia also gratefully acknowledges the grant FPU14/04104 received from the program 'Ayudas para la Formación de Profesorado Universitario (convocatoria 2014)', from Ministerio de Educación, Cultura y Deporte, Gobierno de España.

## Supplementary material

This material contains, at temperature $T = 298.15$ K and pressure $p = 0.1013$ MPa: i) the physical properties and Flory reduction parameters of the pure compounds; and ii) the experimental excess molar volumes, together with the corresponding Flory calculations, for ketone or dialkyl carbonate + alkane systems.

**Table 1**

Excess molar enthalpies, $H_m^E$, at temperature $T$, pressure $p = 0.1013$ MPa and equimolar composition; $X_{12}$, Flory energetic parameter calculated from $H_m^E$ at equimolar composition; $H_{m,int}^E$ and $H_{m,eos}^E$ are the interactional and equation of state contributions; $\sigma_r(H_m^E)$ are the relative standard deviations calculated according to eq. (17).

| System | $X_{12}$/MPa | $H_m^E$/J·mol$^{-1}$ | $H_{m,int}^E$/J·mol$^{-1}$ | $H_{m,eos}^E$/J·mol$^{-1}$ | $\sigma_r(H_m^E)$ | Ref. |
|---|---|---|---|---|---|---|
| ketone + alkane ; $T/K = 298.15$ | | | | | | |
| 2-propanone + hexane (243.15 K) | 73.26 | 1256 | 978 | 278 | 0.183 | [72] |
| 2-propanone + hexane (253.15 K) | 78.82 | 1358 | 1036 | 322 | 0.159 | [72] |
| 2-propanone + hexane (273.15 K) | 83.75 | 1462 | 1066 | 396 | 0.153 | [72] |
| 2-propanone + hexane (293.15 K) | 87.61 | 1554 | 1081 | 473 | 0.151 | [72] |
| 2-propanone + heptane | 91.85 | 1676 | 1176 | 500 | 0.132 | [42] |
| 2-propanone + decane | 102.54 | 1968 | 1456 | 512 | 0.249 | [50] |
| 2-propanone + hexadecane [a] | 116.57 | 2377 | 1866 | 511 | 0.179 | [50] |
| 2-butanone + pentane | 58.85 | 1160 | 805 | 355 | 0.149 | [39] |
| 2-butanone + hexane | 60.69 | 1254 | 881 | 373 | 0.144 | [40] |
| 2-butanone + heptane | 62.71 | 1338 | 955 | 383 | 0.148 | [39] |
| 2-butanone + octane | 64.32 | 1408 | 1021 | 387 | 0.144 | [39] |
| 2-butanone + decane | 67.93 | 1545 | 1150 | 395 | 0.139 | [39] |
| 2-butanone + dodecane | 71.09 | 1661 | 1268 | 393 | 0.216 | [41] |
| 2-butanone + hexadecane | 76.32 | 1863 | 1463 | 400 | 0.214 | [41] |
| 2-pentanone + pentane | 43.92 | 966 | 694 | 272 | 0.151 | [39] |
| 2-pentanone + hexane | 44.48 | 1040 | 748 | 292 | 0.149 | [40] |
| 2-pentanone + heptane | 46.63 | 1135 | 822 | 313 | 0.144 | [39] |
| 2-pentanone + octane | 47.85 | 1202 | 881 | 321 | 0.140 | [39] |
| 2-pentanone + decane | 50.75 | 1335 | 998 | 337 | 0.136 | [39] |
| 3-pentanone + pentane | 42.14 | 920 | 662 | 258 | 0.150 | [39] |
| 3-pentanone + hexane | 43.06 | 999 | 718 | 281 | 0.146 | [40] |
| 3-pentanone + heptane | 44.60 | 1078 | 781 | 297 | 0.140 | [39] |
| 3-pentanone + octane | 45.76 | 1141 | 836 | 305 | 0.136 | [39] |
| 3-pentanone + decane | 48.35 | 1262 | 944 | 318 | 0.126 | [39] |
| 2-hexanone + hexane | 36.93 | 949 | 694 | 255 | 0.179 | [43] |
| 2-hexanone + heptane | 39.16 | 1055 | 773 | 282 | 0.158 | [43] |
| 2-hexanone + octane | 40.54 | 1132 | 835 | 297 | 0.151 | [43] |
| 2-hexanone + nonane | 41.70 | 1198 | 894 | 304 | 0.134 | [43] |
| 2-hexanone + decane | 43.12 | 1268 | 951 | 317 | 0.137 | [43] |
| 2-heptanone + heptane | 30.26 | 886 | 662 | 224 | 0.145 | [44] |
| 4-heptanone + heptane | 27.78 | 812 | 609 | 203 | 0.138 | [44] |
| 2-octanone + dodecane | 29.06 | 1068 | 822 | 246 | 0.144 | [59] |
| 2-octanone + tetradecane | 30.29 | 1151 | 901 | 250 | 0.149 | [59] |
| 2-octanone + hexadecane | 32.03 | 1256 | 984 | 272 | 0.133 | [59] |
| 2-decanone + dodecane | 22.12 | 937 | 725 | 212 | 0.157 | [73] |
| 2-decanone + tetradecane | 23.16 | 1020 | 799 | 221 | 0.164 | [73] |
| 2-decanone + hexadecane | 24.71 | 1126 | 883 | 243 | 0.151 | [73] |
| 2-propanone + $C_6H_{12}$ | 94.19 | 1580 | 1116 | 464 | 0.139 | [74] |
| 2-butanone + $C_6H_{12}$ | 65.93 | 1286 | 923 | 363 | 0.163 | [75] |
| 2-pentanone + $C_6H_{12}$ | 51.80 | 1147 | 834 | 313 | 0.156 | [45] |
| 3-pentanone + $C_6H_{12}$ | 48.57 | 1068 | 777 | 291 | 0.121 | [45] |
| 2-hexanone + $C_6H_{12}$ | 42.79 | 1044 | 767 | 277 | 0.126 | [45] |
| *c*-pentanone + hexane | 56.94 | 1165 | 869 | 296 | 0.177 | [53] |
| *c*-pentanone + heptane | 58.64 | 1256 | 936 | 320 | 0.165 | [53] |
| *c*-pentanone + octane | 61.01 | 1353 | 1012 | 341 | 0.176 | [53] |
| *c*-pentanone + decane | 62.82 | 1465 | 1108 | 357 | 0.194 | [53] |
| *c*-pentanone + dodecane | 66.40 | 1605 | 1232 | 373 | 0.190 | [53] |
| *c*-pentanone + hexadecane | 85.42 | 2174 | 1695 | 479 | 0.054 | [76] |
| *c*-hexanone + hexane | 45.78 | 1035 | 794 | 241 | 0.118 | [53] |
| *c*-hexanone + heptane | 47.77 | 1139 | 867 | 272 | 0.145 | [53] |
| *c*-hexanone + octane | 51.72 | 1285 | 976 | 309 | 0.149 | [53] |
| *c*-hexanone + decane | 52.49 | 1381 | 1055 | 326 | 0.149 | [53] |
| *c*-hexanone + dodecane | 54.69 | 1500 | 1158 | 342 | 0.149 | [53] |
| *c*-hexanone + hexadecane | 67.49 | 1962 | 1534 | 428 | 0.107 | [76] |
| *c*-pentanone + $C_6H_{12}$ | 57.76 | 1132 | 847 | 285 | 0.153 | [76] |
| *c*-hexanone + $C_6H_{12}$ | 43.31 | 939 | 718 | 221 | 0.139 | [76] |
| acetophenone + pentane | 58.69 | 1339 | 1056 | 283 | 0.132 | [58] |
| acetophenone + hexane | 58.39 | 1437 | 1115 | 322 | 0.132 | [77] |
| acetophenone + hexane (328.15 K) | 61.31 | 1515 | 1131 | 384 | 0.139 | [78] |
| acetophenone + heptane | 57.54 | 1493 | 1152 | 341 | 0.135 | [58] |
| acetophenone + heptane (328.15 K) | 61.88 | 1619 | 1197 | 422 | 0.129 | [78] |
| acetophenone + heptane (348.15 K) | 63.61 | 1676 | 1203 | 473 | 0.120 | [78] |
| acetophenone + octane | 57.03 | 1543 | 1190 | 353 | 0.142 | [58] |
| acetophenone + decane | 56.24 | 1620 | 1253 | 367 | 0.151 | [58] |
| acetophenone + $C_6H_{12}$ | 56.91 | 1338 | 1036 | 302 | 0.148 | [78] |
| acetophenone + $C_6H_{12}$ (328.15 K) | 65.24 | 1550 | 1148 | 402 | 0.131 | [78] |
| acetophenone + $C_6H_{12}$ (348.15 K) | 66.78 | 1598 | 1149 | 449 | 0.124 | [78] |
| aldehyde + alkane ; $T / K = 298.15$ | | | | | | |
| ethanal + hexane | 115.99 | 1669 | 1126 | 543 | 0.184 | [24] |
| ethanal + heptane | 122.9 | 1796 | 1251 | 545 | 0.191 | [24] |
| ethanal + dodecane | 144.78 | 2231 | 1705 | 526 | 0.175 | [24] |
| ethanal + hexadecane | 156.46 | 2487 | 1963 | 524 | 0.152 | [24] |

| Mixture | | | | | | |
|---|---|---|---|---|---|---|
| propanal + hexane | 76.09 | 1343 | 926 | 417 | 0.226 | [24] |
| propanal + heptane | 80.57 | 1457 | 1027 | 430 | 0.330 | [24] |
| propanal + dodecane | 96.16 | 1863 | 1424 | 439 | 0.155 | [24] |
| propanal + hexadecane | 101.92 | 2045 | 1615 | 430 | 0.182 | [24] |
| butanal + hexane | 54.37 | 1126 | 794 | 332 | 0.181 | [24] |
| butanal + heptane | 56.87 | 1218 | 871 | 347 | 0.182 | [24] |
| butanal + dodecane | 67.30 | 1576 | 1205 | 371 | 0.183 | [24] |
| butanal + hexadecane | 73.25 | 1792 | 1409 | 383 | 0.163 | [24] |
| benzaldehyde + hexane | 55.88 | 1241 | 974 | 267 | 0.236 | [60] |
| benzaldehyde + heptane | 57.98 | 1360 | 1057 | 303 | 0.190 | [60] |
| linear organic carbonate + $n$-alkane ; $T/\text{K} = 298.15$ | | | | | | |
| dimethyl carbonate + heptane | 96.93 | 1988 | 1419 | 569 | 0.112 | [61] |
| dimethyl carbonate + decane | 100.57 | 2205 | 1631 | 574 | 0.112 | [61] |
| diethyl carbonate + heptane | 50.26 | 1328 | 951 | 377 | 0.111 | [62] |
| diethyl carbonate + decane | 54.01 | 1536 | 1142 | 394 | 0.095 | [62] |
| ketone + ketone ; $T/\text{K} = 303.15$ | | | | | | |
| 2-propanone + 2-butanone | 1.95 | 29 | 22 | 7 | 0.035 | [79] |
| 2-propanone + 2-pentanone | 6.65 | 103 | 80 | 23 | 0.047 | [79] |
| 2-propanone + 2-heptanone | 16.27 | 269 | 215 | 54 | 0.020 | [79] |
| 2-propanone + 2-octanone | 21.49 | 367 | 292 | 75 | 0.033 | [79] |
| 2-propanone + 2-undecanone | 30.99 | 560 | 458 | 102 | 0.047 | [79] |
| 2-butanone + 2-pentanone | 0.99 | 17 | 14 | 3 | 0.061 | [79] |
| 2-butanone + 2-heptanone | 5.16 | 97 | 81 | 16 | 0.043 | [79] |
| 2-butanone + 2-octanone | 8.27 | 163 | 133 | 30 | 0.032 | [79] |
| 2-butanone + 2-undecanone | 16.64 | 354 | 292 | 62 | 0.032 | [79] |
| 2-pentanone + 2-heptanone | 1.77 | 38 | 31 | 7 | 0.054 | [79] |
| 2-pentanone + 2-octanone | 3.51 | 81 | 66 | 15 | 0.031 | [79] |
| 2-pentanone + 2-undecanone | 8.82 | 218 | 179 | 39 | 0.031 | [79] |
| ketone + linear organic carbonate ; $T/\text{K} = 298.15$ | | | | | | |
| 2-propanone + dimethyl carbonate | 12.51 | 187 | 139 | 48 | 0.207 | [80] |
| 2-propanone + diethyl carbonate | 13.74 | 232 | 169 | 63 | 0.378 | [80] |
| 2-butanone + dimethyl carbonate | 15.26 | 274 | 198 | 76 | 0.173 | [80] |
| 2-butanone + diethyl carbonate | 7.17 | 144 | 103 | 41 | 0.512 | [80] |
| 2-pentanone + dimethyl carbonate | 18.56 | 376 | 275 | 101 | 0.200 | [80] |
| 2-pentanone + diethyl carbonate | 6.82 | 154 | 113 | 41 | 0.364 | [80] |
| 2-hexanone + dimethyl carbonate | 23.43 | 519 | 383 | 136 | 0.251 | [80] |
| 2-hexanone + diethyl carbonate | 8.67 | 214 | 161 | 53 | 0.219 | [80] |
| 2-octanone + dimethyl carbonate | 29.11 | 742 | 562 | 180 | 0.120 | [80] |
| 2-octanone + diethyl carbonate | 12.43 | 351 | 275 | 76 | 0.098 | [80] |
| 2-undecanone + dimethyl carbonate | 34.58 | 1028 | 799 | 229 | 0.054 | [80] |
| 2-undecanone + diethyl carbonate | 17.15 | 568 | 458 | 110 | 0.033 | [80] |
| *c*-hexanone + dimethyl carbonate | 25.81 | 502 | 392 | 110 | 0.052 | [81] |
| *c*-hexanone + diethyl carbonate | 11.91 | 244 | 203 | 41 | 0.050 | [81] |

[a] There is a partial immiscibility region.

**Table 2**

Standard molar enthalpies of vaporization, $\Delta H_{m,v}$, differences of standard molar enthalpy of vaporization with respect to the homomorphic hydrocarbon at 298.15 K, $\Delta\Delta H_{m,v}$ (eq. (19)), dipole moments, $\mu$, and effective dipole moments, $\bar{\mu}$ (eq. (20)), at $T = 298.15$ K and $p = 0.1013$ MPa of some pure compounds.

| Compound | $\Delta H_{m,v}$/kJ·mol$^{-1}$ | $\Delta\Delta H_{m,v}$/kJ·mol$^{-1}$ [a] | $\mu$/D | $\bar{\mu}$ |
|---|---|---|---|---|
| 2-propanone | 30.99 [54] | 16.20 | 2.88 [55] | 1.281 |
| 2-butanone | 34.79 [54] | 13.17 | 2.779 [55] | 1.12 |
| 2-pentanone | 38.40 [54] | 11.97 | 2.70 [55] | 0.996 |
| 3-pentanone | 38.52 [54] | 12.09 | 2.82 [55] | 1.046 |
| 2-hexanone | 42.97 [56] | 11.41 | 2.66 [55] | 0.913 |
| 2-heptanone | 47.24 [54] | 10.67 | 2.59 [55] | 0.835 |
| 4-heptanone | 47.8 [82] | 11.23 | | |
| 2-octanone | 51.8 [83] | 10.31 | 2.70 [55] | 0.823 |
| c-pentanone | 42.72 [54] | 14.20 | 3.3 [55] | 1.337 |
| c-hexanone | 45.06 [54] | 12.05 | 3.246 [55] | 1.216 |
| acetophenone | 53.39 [83] | 11.15 | 3.02 [55] | 1.066 |
| ethanal | 25.47 [54] | 20.31 | 2.750 [55] | 1.392 |
| propanal | 29.62 [54] | 14.83 | 2.72 [55] | 1.214 |
| butanal | 33.68 [83] | 12.06 | 2.72 [55] | 1.093 |
| benzaldehyde | 50.30 [83] | 12.29 | 3.0 [55] | 1.136 |
| dimethyl carbonate | 38.0 [84] | 16.38 [b] | 0.94 [65] | 0.391 |
| diethyl carbonate | 43.60 [54] | 12.04 [b] | 0.90 [65] | 0.312 |

[a] Values of $\Delta H_{m,v}$ of the corresponding homomorphic hydrocarbon were taken from the recommended values in ref. [54]. [b] Relative to the *n*-alkane with one more C atom.

**Table 3**

Contribution to the excess molar enthalpy of interactions between unlike molecules, $\Delta H_m^{A\text{-}B}$ (eq. (21)), of ketone + ketone or + linear organic carbonate systems at $T = 298.15$ K and $p = 0.1013$ MPa.

| System | $\Delta H_m^{A\text{-}B}$ /J·mol$^{-1}$ [a] |
|---|---|
| ketone + ketone | |
| 2-propanone + 2-butanone | –2985 |
| 2-propanone + 2-pentanone | –2708 |
| 2-propanone + 2-heptanone | –2293 |
| 2-butanone + 2-pentanone | –2456 |
| 2-butanone + 2-heptanone | –2127 |
| 2-pentanone + 2-heptanone | –1983 |
| ketone + linear organic carbonate | |
| 2-propanone + dimethyl carbonate | –3477 |
| 2-propanone + diethyl carbonate | –2772 |
| 2-butanone + dimethyl carbonate | –3052 |
| 2-butanone + diethyl carbonate | –2522 |
| 2-pentanone + dimethyl carbonate | –2747 |
| 2-pentanone + diethyl carbonate | –2309 |
| 2-hexanone + dimethyl carbonate | –2524 |
| 2-hexanone + diethyl carbonate | –2169 |
| c-hexanone + dimethyl carbonate | –2625 |
| c-hexanone + diethyl carbonate | –2223 |

[a] For references of $H_m^E$(ketone + $n$-heptane), see **Table 1**. The values of $H_m^E$(linear organic carbonate + $n$-heptane) were taken from refs. [61, 62].

**Fig. 1**

Mean relative standard deviations of the Flory model, $\langle \sigma_r(H_m^E) \rangle$ (eq. (18)), of several homologous series of the binary mixtures 2-alkanone + $n$-alkane, or + 2-alkanone, or + dialkyl carbonate.

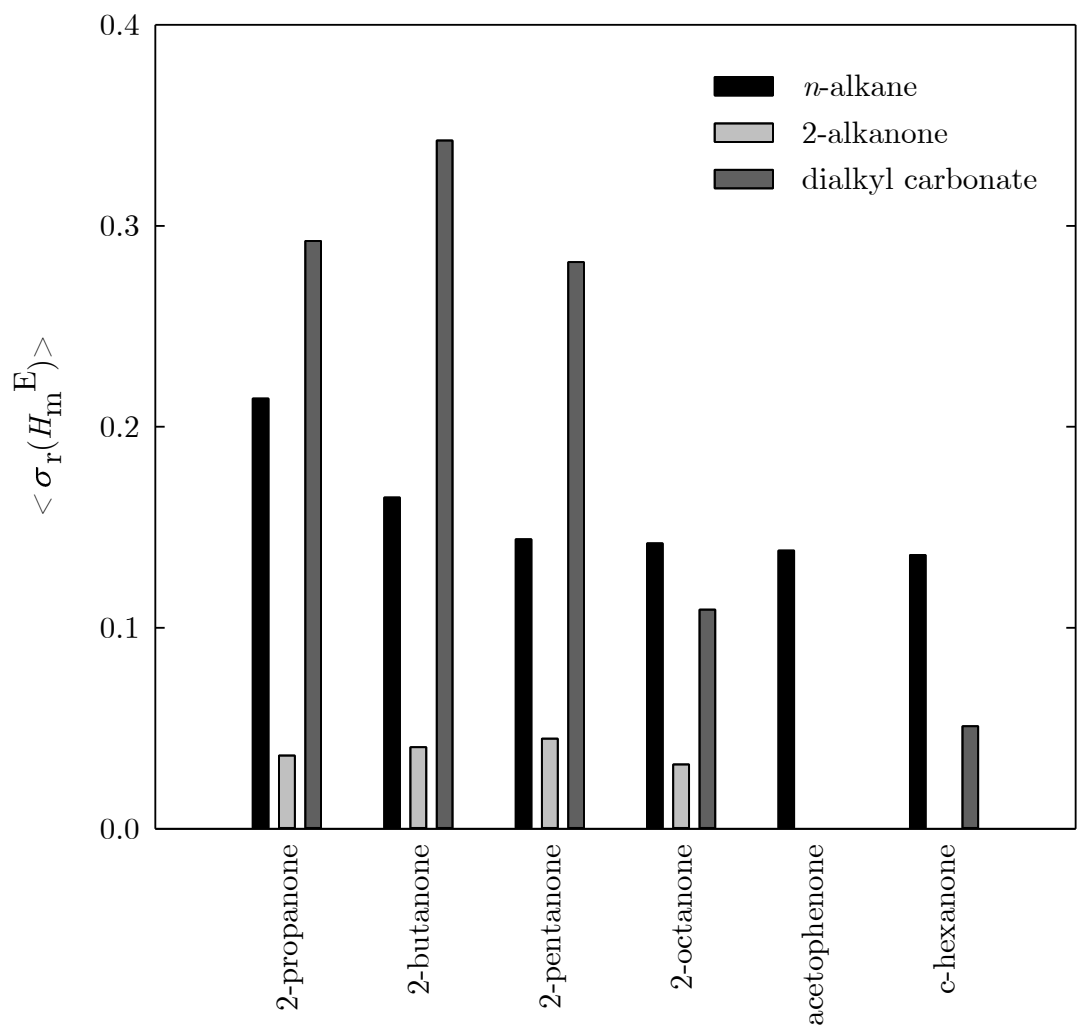

**Fig. 2**

Mean relative standard deviations of the Flory model, $\langle \sigma_r(H_m^E) \rangle$ (eq. (18)), of several homologous series of binary systems of the form ketone or alkanal + *n*-alkane (2 C atoms, ethanal; 3 C atoms, 2-propanone or propanal; 4 C atoms, 2-butanone or butanal; aromatic, acetophenone or benzaldehyde.

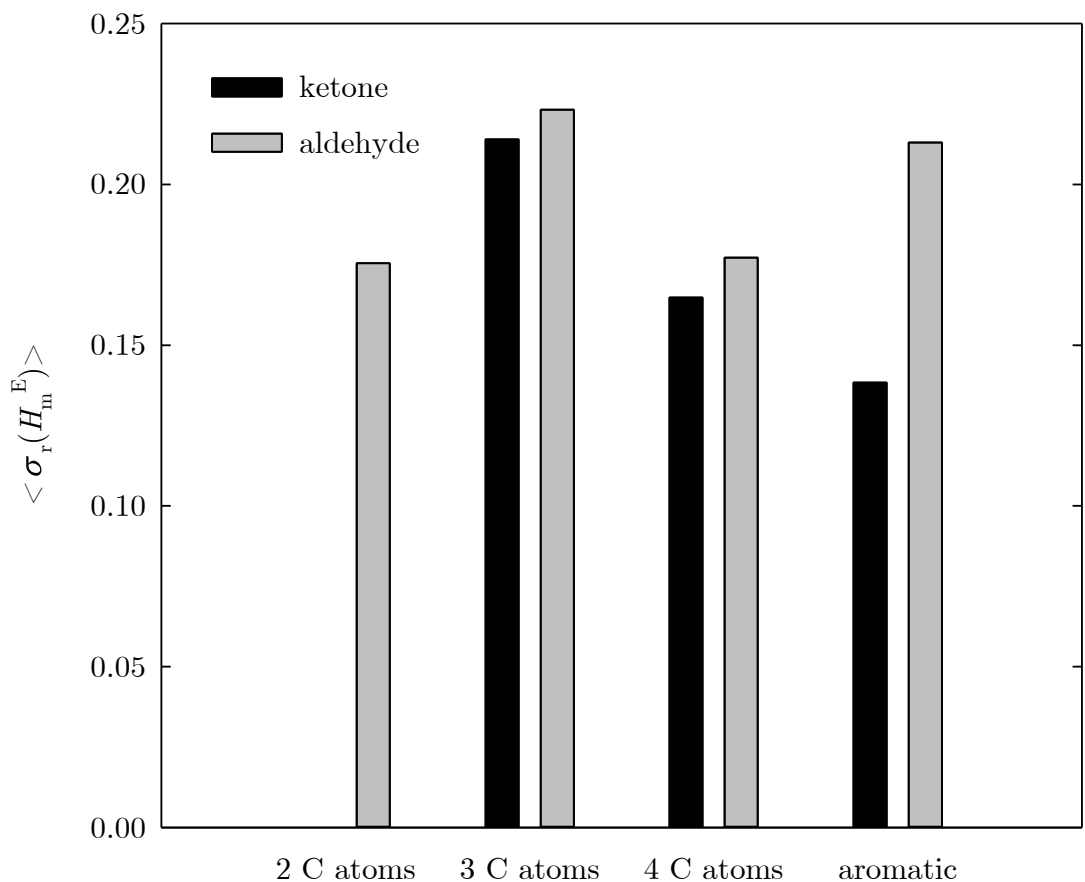

**Fig. 3**

Representation of $X'_{12}(x_1)/X_{12}$ (eq. (13)) for several systems ketone or aldehyde + heptane as functions of the mole fraction of component 1, $x_1$. Symbols: 2-propanone (●), 2-butanone (○), 2-pentanone (▼), 2-hexanone (△), 2-heptanone (■), cyclopentanone (□), cyclohexanone (◆), ethanal (◇), propanal (▲), and butanal (▽).

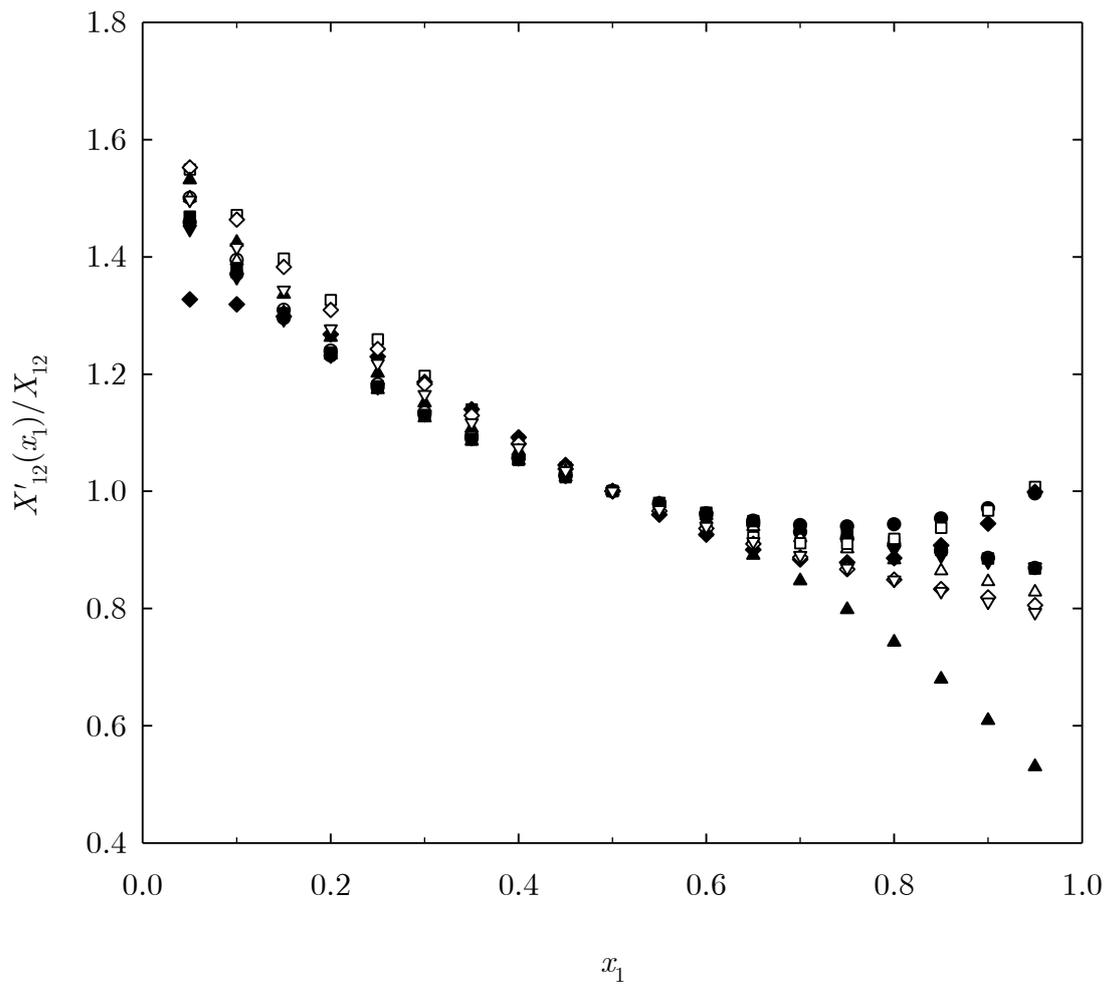

**Fig. 4**

Representation of $X'_{12}(x_1)/X_{12}$ (eq. (13)) for several systems ether + heptane at 298.15 K as functions of the mole fraction of component 1, $x_1$. Symbols: dipropyl ether (●), 2,5-dioxahexane (○), 2,5,8-trioxanonane (▼), 2,5,8,11,14-pentaoxapentadecane (△).

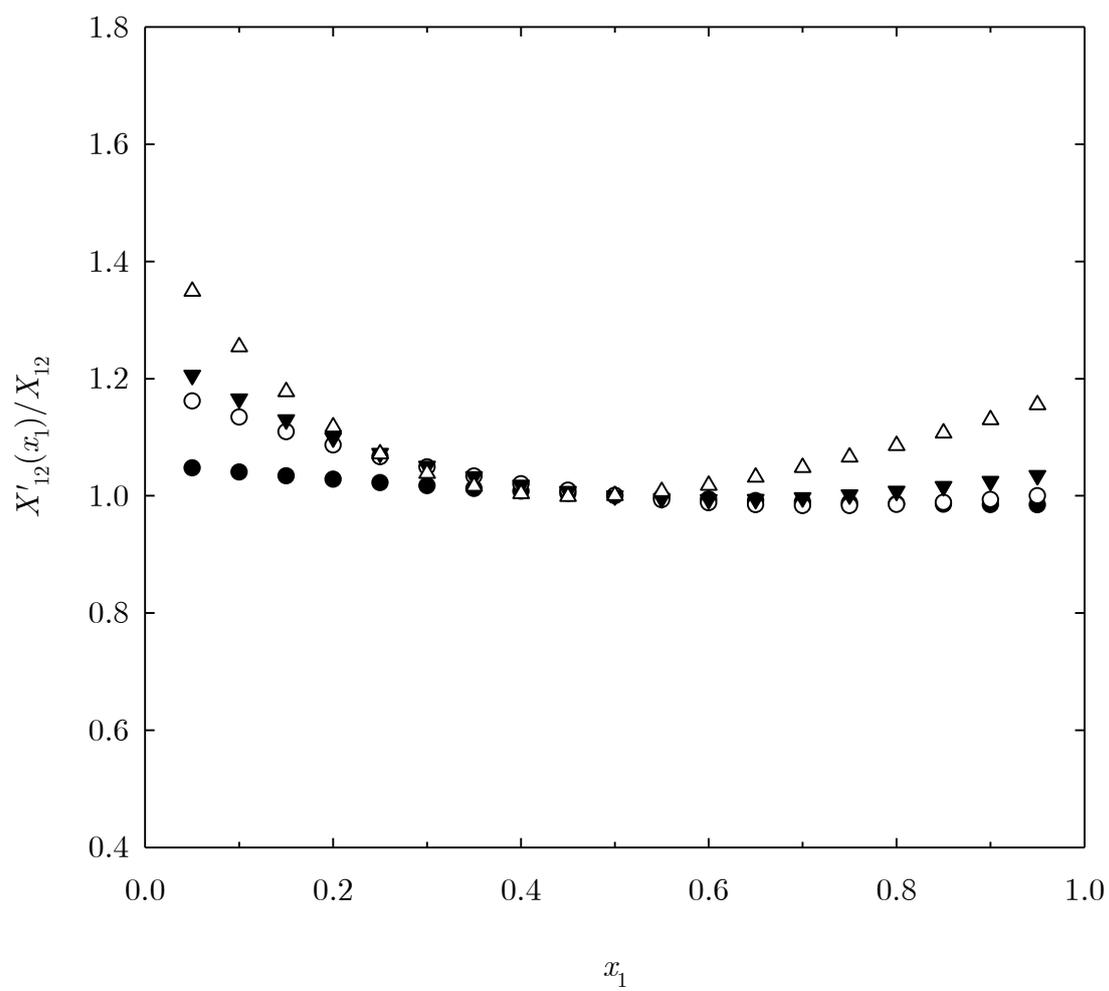

**Fig. 5**

Contributions to the excess molar volume, $V_m^E$, according to the Prigogine-Flory-Patterson model, represented in terms of the difference between the reduced volumes of the components 1 and 2, $\bar{v}_1 - \bar{v}_2$. Solid lines, 2-alkanone + heptane systems (the numbers indicate the carbon atoms of the 2-alkanone). Dashed lines, cyclohexanone + $n$-alkane mixtures ($n$-C$_6$, hexane; $n$-C$_7$, heptane). Full symbols: interactional contribution (●), $p^*$ term (▼), total $V_m^E$ from the model (■), experimental values (◆).

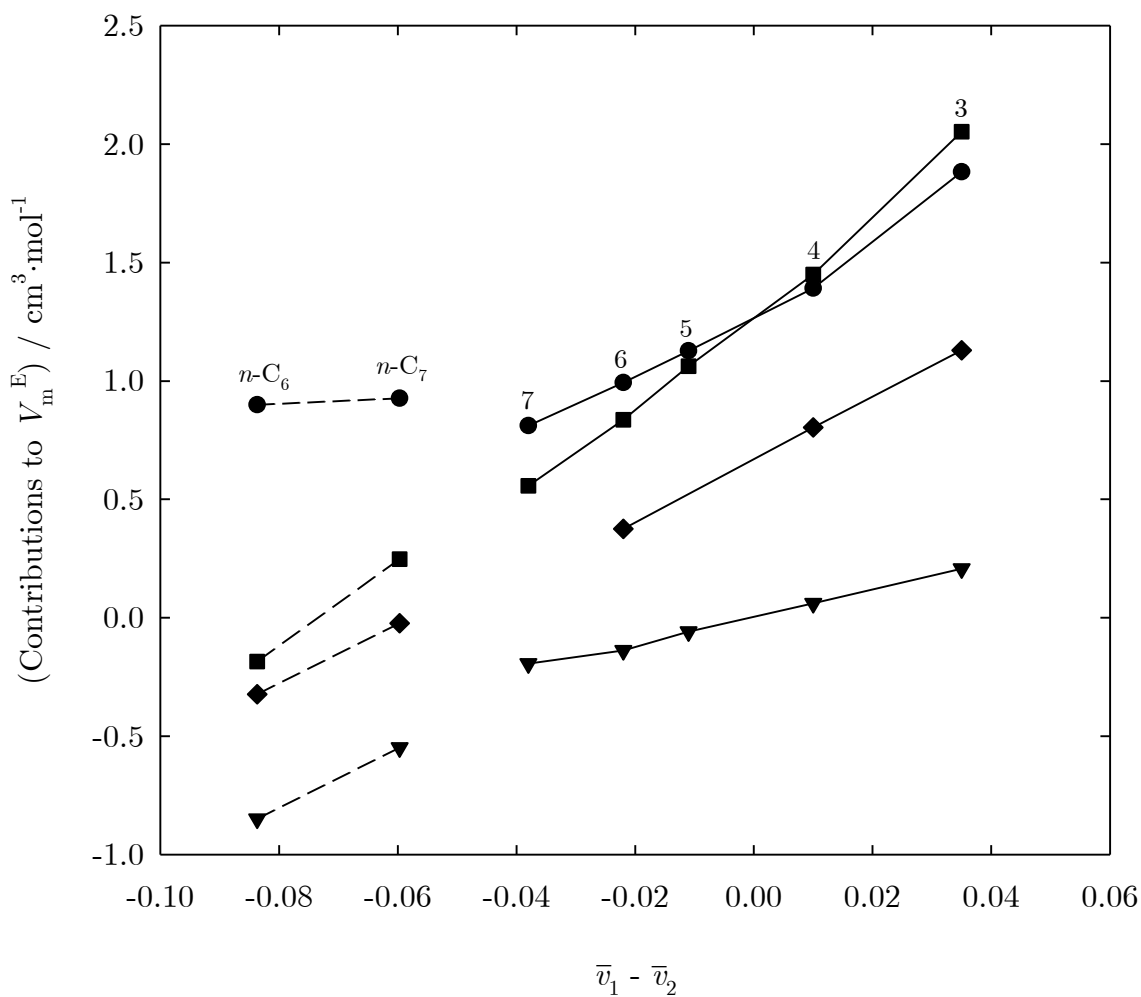

**Fig. 6**

Excess molar enthalpies, $H_m^E$, of 2-alkanone + heptane systems as functions of the mole fraction of component 1, $x_1$. Full symbols, smoothed experimental results: 2-propanone (●), 2-butanone (▼), 2-pentanone (■), 2-hexanone (◆), 2-heptanone (▲). Solid lines, calculations from the Flory model.

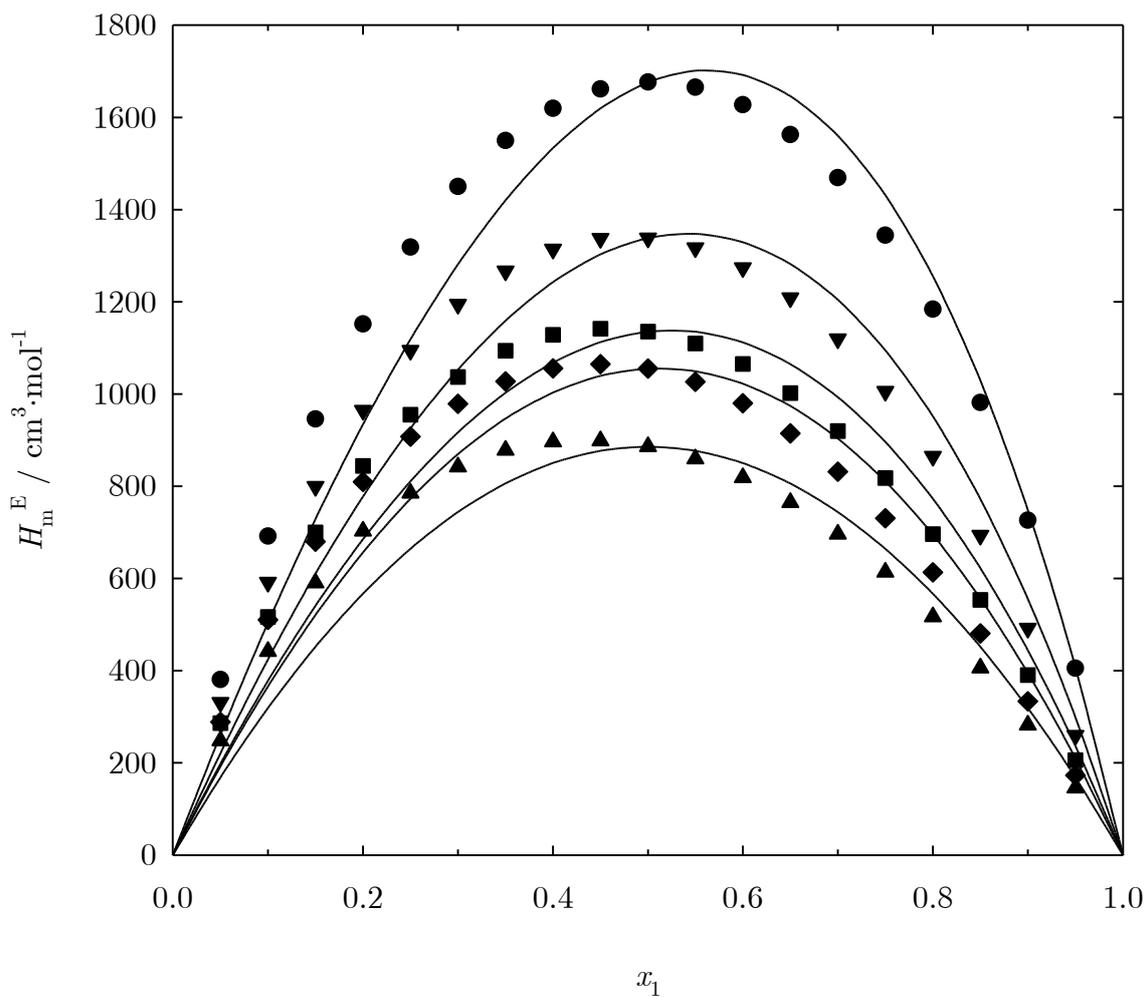

**Fig. 7**

Excess molar enthalpies, $H_m^E$, of 2-propanone + $n$-alkanone systems, as functions of the mole fraction of component 1, $x_1$. Full symbols, smoothed experimental results: 2-butanone (●), 2-pentanone (▼), 2-heptanone (■), 2-octanone (◆), 2-undecanone (▲). Solid lines, calculations from the Flory model.

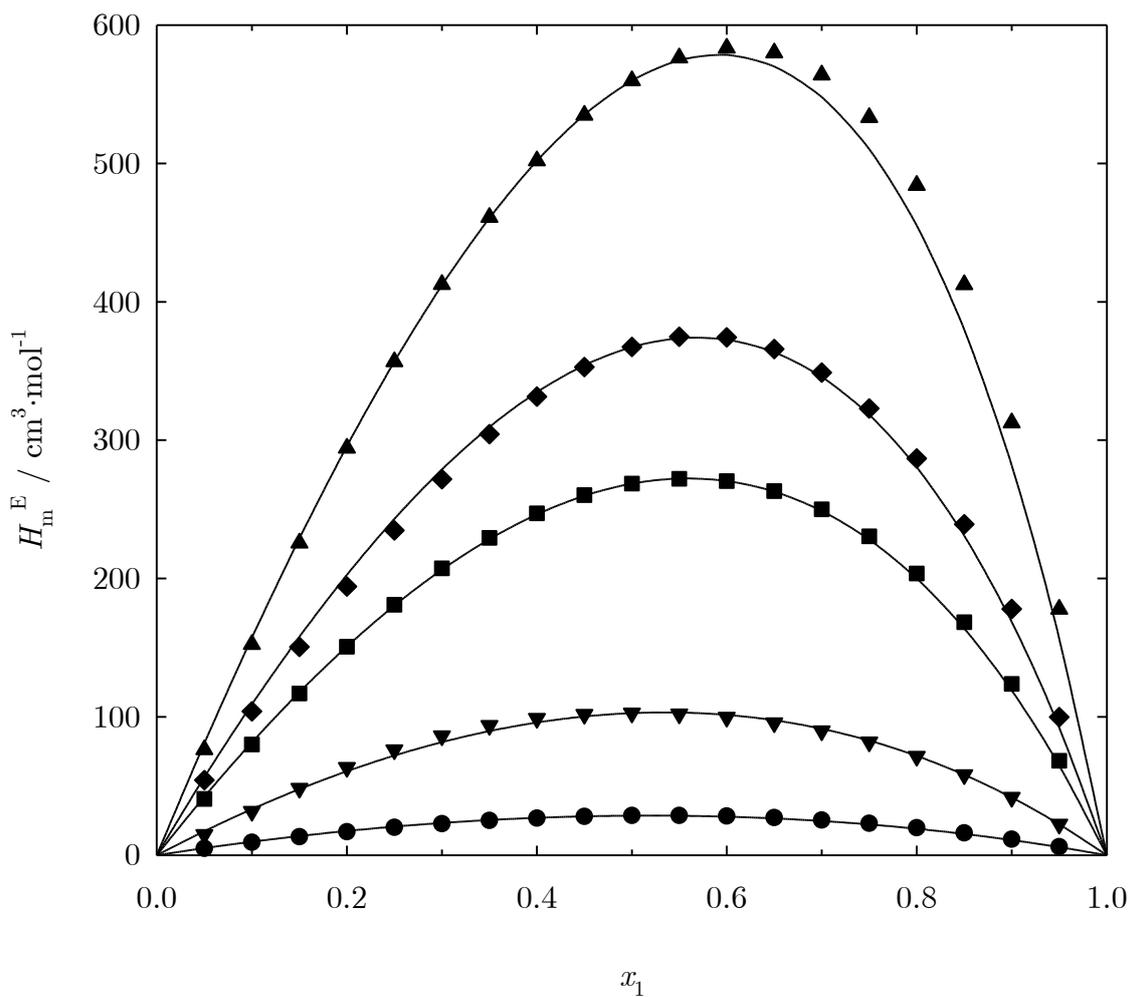

**Fig. 8**

Excess molar enthalpies, $H_m^E$, of *n*-alkanone + dimethyl carbonate systems, as functions of the mole fraction of component 1, $x_1$. Full symbols, smoothed experimental results: 2-undecanone (●), cyclohexanone (▼). Solid lines, calculations from the Flory model.

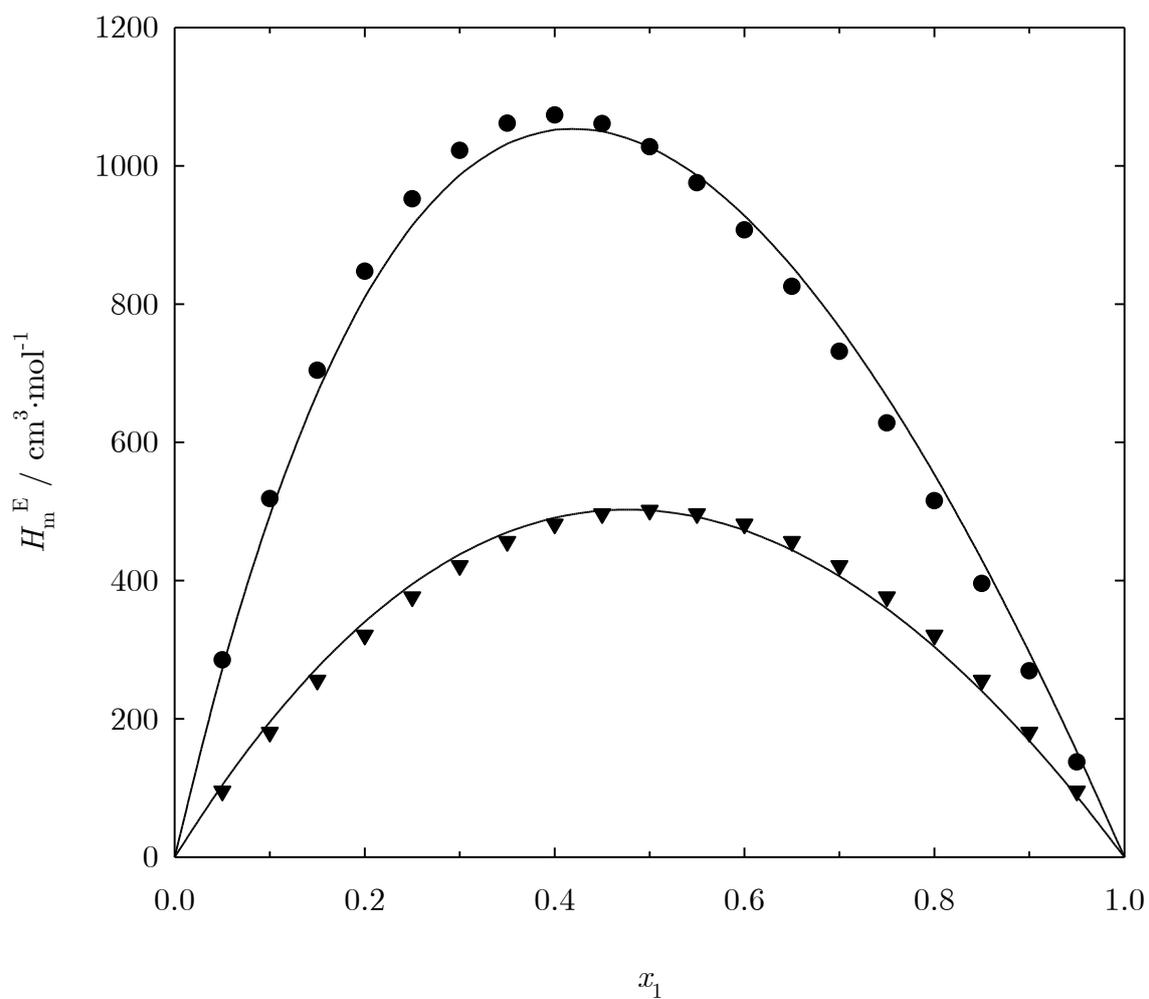

**Supplementary material**

# ORIENTATIONAL EFFECTS IN ALKANONE, ALKANAL OR DIALKYL CARBONATE + ALKANE MIXTURES AND IN ALKANONE + ALKANONE + OR + DIALKYL CARBONATE SYSTEMS.

FERNANDO HEVIA[1], JUAN ANTONIO GONZALEZ[1]*, CRISTINA ALONSO-TRISTÁN[2], ISAÍAS GARCÍA DE LA FUENTE[1] AND LUIS FELIPE SANZ[1]

[1] G.E.T.E.F., Departamento de Física Aplicada, Facultad de Ciencias, Universidad de Valladolid. Paseo de Belén, 7, 47011 Valladolid, Spain.

[2] Departamento de Ingeniería Electromecánica. Escuela Politécnica Superior, Universidad de Burgos. Avda. Cantabria s/n. 09006 Burgos, Spain.

*CORRESPONDING AUTHOR, e-mail: jagl@termo.uva.es; Fax: +34-983-423136; Tel: +34-983-423757

**Table S1**

Physical properties and Flory reduction parameters of the pure compounds at temperature $T = 298.15$ K and pressure $p = 0.1013$ MPa. $V_m$, molar volume; $\alpha_p$, isobaric coefficient of thermal expansion; $\kappa_T$, coefficient of isothermal compressibility; $V_m^*$, reduction molar volume; and $p^*$, reduction pressure.

| Compound | $V_m$/cm$^3$·mol$^{-1}$ | $\alpha_p$/10$^{-3}$K$^{-1}$ | $\kappa_T$/TPa$^{-1}$ | $V_m^*$/cm$^3$·mol$^{-1}$ | $p^*$/MPa |
|---|---|---|---|---|---|
| $n$-pentane | 116.11[1] | 1.61[1] | 2180[1] | 85.33 | 407.5 |
| $n$-hexane | 131.57[2] | 1.387[2] | 1794[2] | 99.52 | 402.7 |
| $n$-heptane | 147.45[3] | 1.256[3] | 1461[3] | 113.60 | 431.7 |
| $n$-octane | 163.52[3] | 1.164[3] | 1302.4[3] | 127.70 | 436.8 |
| $n$-nonane | 179.69[1] | 1.0844[1] | 1177[1] | 142.07 | 439.3 |
| $n$-decane | 195.90[3] | 1.051[3] | 1110[3] | 155.71 | 446.7 |
| $n$-dodecane | 228.47[3] | 0.96[3] | 988[3] | 184.33 | 444.9 |
| $n$-tetradecane | 261.09[3] | 0.886[4] | 872[4] | 213.33 | 453.6 |
| $n$-hexadecane | 294.04[3] | 0.883[2] | 862[2] | 240.38 | 456.8 |
| $c$-hexane | 108.73[1] | 1.22[1] | 1129.2[1] | 84.21 | 536.9 |
| 2-propanone | 74.00[5] | 1.45[5] | 1317.5[5] | 55.50 | 583.2 |
| 2-butanone | 90.14[6] | 1.31[7] | 1175.9[7] | 68.91 | 568.2 |
| 2-pentanone | 107.46[8] | 1.198[8] | 1098.77[8] | 83.50 | 538.2 |
| 3-pentanone | 106.41[1] | 1.2[9] | 1073[9] | 82.66 | 552.4 |
| 2-hexanone | 124.16[1] | 1.14[10] | 1012[1] | 97.32 | 546.5 |
| 2-heptanone | 140.76[11] | 1.06[11] | 968.9[11] | 111.72 | 517.6 |
| 4-heptanone | 140.79[1] | 1.05[12] | 932[12] | 111.92 | 531.4 |
| 2-octanone | 157.45[1] | 1.03[1] | 899[1] | 125.57 | 536.9 |
| 2-decanone | 190.55[13] | 0.97 [a] | 861.0 [b] | 153.48 | 517.6 |
| 2-undecanone | 207.12[13] | 0.94 [a] | 833.4 [b] | 167.67 | 513.0 |
| $c$-pentanone | 89.06[14] | 1.023 [c] | 725.2 [d] | 71.11 | 659.6 |
| $c$-hexanone | 104.21[1] | 0.955 [e] | 695 [e] | 84.15 | 628.1 |
| acetophenone | 117.35[1] | 0.87[1] | 710.6 [f] | 96.15 | 543.6 |
| ethanal | 57.06[15] | 1.69[1] | 1400 [g] | 41.53 | 679.2 |
| propanal | 73.41[1] | 1.472[1] | 1067 [g] | 54.90 | 735.3 |
| butanal | 90.54[1] | 1.306[1] | 1056 [g] | 69.26 | 630.0 |
| benzaldehyde | 102.03[1] | 0.865[1] | 585.6 [h] | 83.67 | 654.7 |
| dimethyl carbonate | 84.71[16] | 1.2541[16] | 893.1[16] | 65.28 | 704.8 |
| diethyl carbonate | 121.90[16] | 1.2971[16] | 970[16] | 93.36 | 679.5 |

[a] Extrapolated from $\alpha_p$ values of smaller 2-alkanones. [b] Estimated using $\alpha_p/\kappa_T$ values obtained by the Manzini-Crescenzi group contribution method [17]. [c] Calculated from data of ref. [14]. [d] Calculated from data of refs. [14, 18]. [e] Calculated from data of ref. [19]. [f] Calculated from data of refs. [1, 20]. [g] Calculated from data of ref. [21]. [h] Calculated from data of ref. [22].

**Table S2**

Experimental excess molar volumes, $V_{m,exp}^{E}$, and Flory calculations, $V_{m,Flory}^{E}$, at temperature $T = 298.15$ K, pressure $p = 0.1013$ MPa and equimolar composition for ketone or dialkyl carbonate + alkane systems.

| System | $V_{m,exp}^{E}$ /cm$^3$·mol$^{-1}$ | $V_{m,Flory}^{E}$ /cm$^3$·mol$^{-1}$ | Ref. |
|---|---|---|---|
| 2-propanone + heptane | 1.130 | 2.060 | [23] |
| 2-propanone + decane | 1.333 | 2.137 | [24] |
| 2-propanone + hexadecane [a] | 1.422 | 2.101 | [24] |
| 2-butanone + heptane | 0.803 | 1.419 | [25] |
| 2-butanone + octane | 0.866 | 1.505 | [26] |
| 2-butanone + decane | 0.952 | 1.590 | [25] |
| 2-butanone + dodecane | 0.996 | 1.657 | [25] |
| 2-pentanone + octane | 0.695 | 1.153 | [27] |
| 2-pentanone + decane | 0.812 | 1.272 | [28] |
| 3-pentanone + heptane | 0.512 | 0.970 | [29] |
| 2-hexanone + hexane | 0.154 | 0.525 | [30] |
| 2-hexanone + heptane | 0.375 | 0.822 | [30] |
| 2-hexanone + octane | 0.516 | 0.977 | [30] |
| 2-hexanone + nonane | 0.595 | 1.091 | [30] |
| 2-hexanone + decane | 0.643 | 1.154 | [30] |
| 2-butanone + $C_6H_{12}$ | 0.912 | 1.140 | [29] |
| 3-pentanone + $C_6H_{12}$ | 0.763 | 0.893 | [29] |
| $c$-pentanone[b] + hexane | −0.172 | 0.159 | [31] |
| $c$-pentanone + heptane | 0.082 | 0.536 | [31] |
| $c$-pentanone + octane | 0.252 | 0.777 | [31] |
| $c$-pentanone + decane | 0.482 | 1.046 | [31] |
| $c$-pentanone + dodecane | 0.634 | 1.291 | [31] |
| $c$-hexanone[c] + hexane | −0.323 | −0.194 | [31] |
| $c$-hexanone + heptane | −0.023 | 0.235 | [31] |
| $c$-hexanone + octane | 0.173 | 0.526 | [31] |
| $c$-hexanone + decane | 0.442 | 0.803 | [31] |
| $c$-hexanone + dodecane | 0.622 | 1.040 | [31] |
| dimethyl carbonate + heptane | 1.158 | 1.830 | [32] |
| dimethyl carbonate + decane | 1.442 | 2.218 | [32] |
| diethyl carbonate + heptane | 0.736 | 1.275 | [33] |
| diethyl carbonate + decane | 1.063 | 1.771 | [33] |

[a] There is a partial immiscibility region; [b]cyclopentanone; [c]cyclohexanone.

# References for supplementary material